\documentclass[11pt]{article}
\def\beq{\begin{equation}}
\def\eeq{\end{equation}}
\def\beqar{\begin{eqnarray}}
\def\eeqar{\end{eqnarray}}
\def\eps{\epsilon}
\def\pp{\;.}
\def\parxx{\frac\partial{\partial x}}
\def\partt{\frac\partial{\partial t}}

\newcommand\fract[2]{\ensuremath{\textstyle\frac{#1}{#2}}}

\newcommand{\To}{\ensuremath{\mathrel{\hbox to
2em{\rightarrowfill}}}}
\newcommand{\dd}[1]{\ensuremath{\mathop{{\rm d}#1}}}

\def\dertt{\frac{\dd{}}{\dd t}}

\renewcommand{\Im}{\ensuremath{\mathop{{\rm Im}}}}
\renewcommand{\Re}{\ensuremath{\mathop{{\rm Re}}}}
\newcommand\brak[1]{\ensuremath{\bigl| #1\bigr\rangle}}
\newcommand\krab[1]{\ensuremath{\bigl\langle #1\bigr|}}

\begin{document}
\title{A Nonrelativistic Chiral Soliton in One Dimension}
\author{\sc R. Jackiw\thanks{Fax 617-253-8674 USA\quad email:
{\tt jackiw@mitlns.mit.edu}\qquad  MIT-CTP \# 2587}\\[2ex] 
Center for Theoretical Physics\\
Massachusetts Institute of Technology\\
6-320\\
77 Massachusetts Avenue\\
Cambridge, MA 02140-4307 USA\\[4ex]
\emph{Physics Abstracts} classification numbers 71.10.Pm,
05.30.--d\\[4ex] Submitted to
\emph{Journal of Nonlinear Mathematical Physics}\\[2ex]
Dedicated to {\sc W. Fushchych}\\
on the occasion of his sixtieth birthday}

\date{}

\maketitle

\begin{abstract}
\noindent
I analyze the one-dimensional, cubic Schr\"odinger
equation, with nonlinearity constructed from the current density,
rather than, as is usual, from the charge density. A soliton
solution is found, where the soliton moves only in one direction.
Relation to higher-dimensional Chern--Simons theory is indicated. The
theory is quantized and results for the two-body quantum problem
 agree at weak coupling with those coming from a
semiclassical quantization of the soliton.
\end{abstract}

\section{Introduction}
The Schr\"odinger equation in one spatial dimension with cubic
nonlinearity
\beqar 
i\hbar \partt\psi(t,x)
&=& \frac{-\hbar^2}{2m} \frac{\partial^2}{\partial x^2} 
\psi(t,x) - g\rho(t,x)\psi(t,x) \label{eq:1}\\
g>0,\quad \rho&=&\psi^*\psi \pp 
\nonumber
\eeqar
plays a cycle of interrelated roles in mathematical physics.
(Here ${}^*$ signifies complex conjugation.)

Viewed as a nonlinear partial differential equation for the function
$\psi$, it possesses the famous soliton solution
\beqar
\psi_s(t,x) &=& \pm e^{i\frac{mv}\hbar (x-ut)} \frac\hbar{\sqrt{gm}} 
\frac\alpha{\cosh \alpha(x-vt)}\label{eq:2}\\
\alpha^2&=&(m^2 v^2/\hbar^2)(1-2u/v)\pp \nonumber
\eeqar
Moreover, Eq.\,(\ref{eq:1}) is completely integrable and multisoliton
solutions can also be explicitly constructed.
 Note that the soliton described by Eq.\,(\ref{eq:2}) moves with
(group) velocity $v$, which is unrestricted as to sign and magnitude.
(The phase velocity $u$ must be less than $v/2$.) In particular the
soliton can be brought to rest: the theory is Galileo-invariant, so that
any solution to~(\ref{eq:1}) can be mapped into another solution by
the Galileo boost
\beqar
x &\To& x - Vt\nonumber\\
\psi(t,x) &\To& e^{\frac i\hbar mV(x-\frac12 Vt)}
\psi(t,x-Vt)\pp \label{eq:3}
\eeqar
[Galileo transformations for the dynamical system~(\ref{eq:1}) are
accompanied by a phase change of $\psi$ -- a so-called ``1-cocycle".]
Applying the transformation~(\ref{eq:3}) to~(\ref{eq:2}) with $V=-v$
results in
\beq
\psi_s(t,x) \To \pm e^{i\frac{\hbar\alpha^2}{2m} t }
\frac\hbar{\sqrt{gm}} \frac\alpha{\cosh\alpha x}
\label{eq:3a}
\eeq
which also solves (\ref{eq:1}) and describes a soliton at rest.

Alternatively one may view~(\ref{eq:1}) as a quantal Heisenberg
equation  of motion for the quantum field operator $\psi(t,x)$, which
is taken to satisfy the commutation relation
  \beq
  \left[ \psi(t,x_1), \psi^*(t,x_2)\right] = \delta(x_1-x_2)\pp
  \label{eq:4}
  \eeq
(Now ${}^*$ denotes Hermitian conjugation.) Since the
dynamics~(\ref{eq:1}) conserves the number operator $\int\dd x
\rho$, the quantum Hilbert space can be decomposed according to the
(integer) eigenvalues $N$ of $\int\dd x\rho$,
and one finds that the $N$-body wave function
   \beq
    \psi_N(t; x_1, \dots, x_N) \equiv \frac1{\sqrt{N!}}
   \krab0 \psi(t,x_1) \cdots \psi(t,x_N)\brak N
   \label{eq:5}
   \eeq
satisfies an $N$-body Schr\"odinger equation with two-body, pairwise
attractive $\delta$-function interactions. The bound-state spectrum
can be found explicitly and the energy eigenvalue is determined as 
   \beq 
   E_N = \frac{-g^2m}{6\hbar^2} (N^3 - N)\pp
   \label{eq:6}
   \eeq

The cycle of formulas closes: the static solution~(\ref{eq:3a}) can be
quantized semiclassically and the bound energy spectrum obtained in
this way coincides with~(\ref{eq:6}).\cite{first} 
Specifically, the classical energy
of~(\ref{eq:3a})
is $\frac{-g^2m}{6\hbar^2} N^3$, where $N$ is the value of $\int\dd
x\rho$ evaluated on~(\ref{eq:3a}). The semiclassical (first quantum)
correction diminishes $N^3$ by $N$ (so that the bound state energy
vanishes at $N=1$). (Also, the semiclassical phase shift of the
scattering problem reproduces the exact quantal formula for
weak coupling.\cite{second})

In this article, I shall review some properties of another nonlinear
Schr\"odinger equation, which (in a transformed formulation) has
recently arisen in discussions of one-dimensional condensed matter
systems (quantum wires, Hall edge states)\,\cite{third} and which
possesses a remarkable soliton solution, in that the soliton is chiral,
{\it i.e.}, it can move only in one direction.\cite{fourth}

\section{The Equation and Its Soliton Solution}
The equation that we consider differs from~(\ref{eq:1}) in that the
nonlinearity involves the current density $j$
   \beq
   j = \frac\hbar{m} \Im  \Bigl(\psi^*\parxx\psi \Bigr)
   \label{eq:7}
   \eeq
rather than charge density~$\rho$:
   \beq
   i\hbar\partt \psi(t,x) =
   -\frac{\hbar^2}{2m} \frac{\partial^2}{\partial x^2} \psi(t,x) - \hbar
   \lambda j(t,x)\psi(t,x)\pp
   \label{eq:8}
   \eeq
Note that $j$ can be substituted for $\rho$ only in one spatial
dimension, where the ``vector" nature of $j$ is absent. The current and
charge densities are linked by the continuity equation, which is valid
both for~(\ref{eq:1}) and~(\ref{eq:8}).
   \beq
   \frac{\partial}{\partial t}\rho + \frac{\partial}{\partial x} j = 0\pp
   \label{eq:9}
   \eeq
When $\psi$ is decomposed into modulus and phase
   \beq 
   \psi = \sqrt\rho e^{i\theta}
   \label{eq:10a}
   \eeq
one sees that
   \beq 
   j =\frac\hbar m \rho\frac{\partial}{\partial x}\theta
   \label{eq:9b}
   \eeq
and (\ref{eq:8}) takes the form of the usual nonlinear Schr\"odinger
equation~(\ref{eq:1}),
  \beq
  i\hbar\partt\psi(t,x) = 
  \frac{-\hbar^2}{2m}\frac{\partial^2}{\partial x^2} \psi(t,x) -
  g(t,x)\rho(t,x)\psi(t,x)
  \label{eq:10}
  \eeq
but with coupling strength 
  \beq
  g(t,x) = \hbar^2 \frac{\lambda}{m}\parxx \theta(t,x)
  \label{eq:11}
  \eeq
modulated by the spatial variation of the phase. 
There is no \emph{a priori} restriction on the sign of $\lambda$, but
for definiteness we shall take it to be positive, $\lambda>0$.

To find the one-soliton solution $\psi_s$, we take the phase to be as in
a propagating wave
  \beq 
  \psi_s = e^{-i(\omega t-kx)}\sqrt \rho
  \label{eq:12a}
  \eeq
so that 
  \beq
  j=v\rho,\qquad v\equiv \frac{\hbar k}m
  \label{eq:12b}
  \eeq
and our equation~(\ref{eq:8}) becomes identical to~(\ref{eq:1}), with
  \beq
  g=\hbar \lambda v\pp
  \label{eq:13a}
  \eeq
The soliton solution exists provided $g>0$; this requires 
  \beq
  v>0 \pp
  \label{eq:13b}
  \eeq 
The soliton can only move to the right; it is chiral. (Had we taken
$\lambda<0$, then the soliton could move only to the left.) The profile
that solves (\ref{eq:8}) is found from (\ref{eq:2}) and  (\ref{eq:13a}) 
to be
  \beq
  \psi_s(t,x) = \pm e^{i\frac{mv}\hbar (x-ut)}
\sqrt{\frac\hbar{\lambda
  mv}}\frac{\alpha}{\cosh \alpha(x-vt)}
  \label{eq:14}
  \eeq
where $u \equiv \omega/k$. The presence of $1/\sqrt v$ in the
amplitude reinforces the statement that the velocity must be positive;
in particular, the soliton cannot be brought to rest. Evidently, the
dynamics (\ref{eq:8}) is \emph{not} Galileo-invariant; the velocity of
the soliton cannot be arbitrarily reduced. This fact will be seen
clearly in the next section, where the action, Lagrangian, and other
dynamical quantities relevant to (\ref{eq:8}) are discussed.

\section{Action Principle, Symmetries, and 
 Constants of Motion}
Eq.\,(\ref{eq:8}) may be obtained from an action principle, which also
is useful for identifying the Hamiltonian (energy), momentum, and
other constants of motion.

Consider the Lagrange density
\beq
{\cal L} = i\hbar\Psi^* \partt \Psi
-\frac{\hbar^2}{2m} 
\left|
   \Bigl(\frac\partial{\partial x} - i\frac\lambda2 \rho\Bigr)\Psi
\right|^2\pp
\label{eq:15}
\eeq
The Euler--Lagrange equation reads
  \beq
  i\hbar\frac\partial{\partial t} \Psi = -\frac{\hbar^2}{2m} 
  \Bigl(\frac\partial{\partial x} - i\frac\lambda2 \rho\Bigr)^2\Psi
  -\frac{\hbar\lambda}{2} j \Psi
  \label{eq:16}
  \eeq
where
  \beqar
  \rho &=& \Psi^*\Psi \label{eq:17a}\\
  j &=& \frac\hbar{m}\Im
\biggl(\Psi^*\Bigl(\parxx - i \frac\lambda2\rho\Bigr)\Psi\biggr)
  \label{eq:17b}
  \eeqar
and the two are linked by the continuity equation~(\ref{eq:9}). 
Next, we redefine the $\Psi$ field by
  \beq
  \Psi(t,x) = e^{i\frac\lambda2 \int^x\dd y \rho(t,y)} \psi(t,x)
  \label{eq:18}
  \eeq
(the lower limit on the integral is immaterial -- it affects only the
phase of $\psi$) so that
   \beq
   i\hbar\frac\partial{\partial t}\psi(t,x) - \frac{\hbar\lambda}2
   \int^x
   \dd y \partt \rho(t,y)\psi(t,x) 
    =  -\frac{\hbar^2}{2m} \frac{\partial^2}{\partial x^2} \psi(t,x)
   -\frac{\hbar\lambda}2 j(t,x) \psi(t,x)\pp
   \label{eq:19}
   \eeq
But the integral may be evaluated with the help of
(\ref{eq:9}) and transferred to the right side. The resulting equation
is just~(\ref{eq:8}).

Energy is conserved as a consequence of time-translation invariance,
and its form can be deduced with the help of Noether's theorem, or by
inspection from~(\ref{eq:15}). Evidently the Hamiltonian (energy)
density is
  \beq
  {\cal E} = \frac{\hbar^2}{2m} \Bigl|
  \Bigl( \frac{\partial}{\partial x} - i\frac\lambda2\rho \Bigr)
  \Psi\Bigr|^2 = \frac{\hbar^2}{2m}\, \Bigl|\parxx \psi\Bigr|^2 \pp
  \label{eq:20}
  \eeq
Similarly, 
space translation invariance ensures momentum conservation, and the
momentum density reads
   \beqar
   {\cal P} &=& \hbar \Im \Bigl( \Psi^*\parxx \Psi\Bigr)\nonumber\\
    &=& mj + \frac{\hbar\lambda}2 \rho^2 \pp
   \label{eq:21}
   \eeqar
These quantities obey continuity equations. With the energy density
$\cal E$  there is associated an energy flux $T^{0x}$
    \beqar
    T^{0 x} &=& \frac{-\hbar^2}{m} \Re
    \Bigl(\partt \Psi^*\Bigl[ \parxx -
      \frac{i\lambda}2 \rho\Bigr]\Psi \Bigr)
    \nonumber\\
    &=& 
    \frac{-\hbar^2}{m} \Re
    \Bigl(\partt \psi^* \parxx \psi \Bigr)
    +\frac{\hbar\lambda}2 j^2
    \label{eq:22}
    \eeqar
and together they satisfy
    \beq
    \frac\partial{\partial t}{\cal E} + \frac\partial{\partial x} T^{0 x}
    =0\pp
    \label{eq:23}
    \eeq
Similarly, with the momentum density $\cal P$, supplemented by the 
momentum flux
  \beqar
   T^{x x} &=& \frac{\hbar^2}m \Bigl| \Bigl(\parxx -
     \frac{i\lambda}2 \rho\Bigr)\Psi\Bigr|^2 - \frac{\hbar^2}{4m}
  \frac{\partial^2}{\partial x^2}\rho \nonumber\\
  &=& \frac{\hbar^2}m
 \Bigl|  \parxx \psi \Bigr|^2
  - \frac{\hbar^2}{4m}
  \frac{\partial^2}{\partial x^2}\rho
  \label{eq:24}
  \eeqar
their continuity equation is verified
  \beq
  \partt {\cal P} + \parxx T^{xx} = 0 \pp
  \label{eq:25}
  \eeq
Therefore, when fields decrease rapidly at spatial infinity, the energy
and momentum are time-independent
  \beqar
   E &=& \int\dd x {\cal E} \qquad \dertt E  =  0 \label{eq:26}\\
   P &=& \int\dd x {\cal P} \qquad\! \dertt P = 0 \pp \label{eq:27}
  \eeqar

Note that the momentum density ${\cal P}$ is not equal to the energy
flux $T^{0x}$, because the theory is not Lorentz-invariant. Also it is not
Galileo-invariant. This is seen in the present context by constructing
the usual Galileo boost  generator
  \beq
  G = tP -m\int \dd{xx} \rho \pp
  \label{eq:28a}
  \eeq
From (\ref{eq:9}) and (\ref{eq:27}) we find
  \beq
  \frac{\dd G}{\dd t} = P + m\int \dd{xx} \parxx j = P - m\int\dd x j \pp
  \label{eq:28b}
  \eeq
But Eq.\,(\ref{eq:21}) shows that $P$ possesses the dynamical
contribution
$\frac{\hbar\lambda}2\int \dd x\rho^2$ in addition to the usual
kinematical term $m\int\dd x j$; therefore it follows that $G$ is not
conserved,
  \beq
  \frac{\dd{}}{\dd t} G = \frac{\hbar\lambda}2\int \dd x \rho^2
  \label{eq:28c}
  \eeq
but always increases in time.

There does exist, however, another symmetry, beyond  time and
space translation invariance and another conserved quantity beyond
$H$ and $P$. The additional symmetry is dilation invariance,
corresponding to the possibility of rescaling space and time:
  \beqar
  t \To at, \quad x\To \sqrt a x, \nonumber\\
  \psi(t,x) \To a^{1/4}\psi(at,\sqrt a x)\pp 
  \label{eq:29}
  \eeqar
Applied to the soliton solution (\ref{eq:14}), the transformation
(\ref{eq:29}) yields another soliton solution with group and phase
velocities $v, u$ rescaled by $\sqrt a$.
In order to construct the constant of motion associated with dilation
invariance, it is useful to ``improve" the energy density and energy
flux by the addition of  terms (``superpotentials") that do not affect
the continuity equation nor the spatial integral defining the energy.
Instead of (\ref{eq:20}) and (\ref{eq:22}), we  use
  \beqar
   {\cal E}_{\rm improved} &=& {\cal E} - \frac{\hbar^2}{8m}
  \frac{\partial^2}{\partial x^2} \rho\label{eq:30}\\
  T^{0x}_{\rm improved} &=& T^{0x} - \frac{\hbar^2}{8m}
  \frac{\partial^2}{\partial x^2} j\pp \label{eq:31}
  \eeqar
Owing to (\ref{eq:9}), we still have
  \beq 
  \partt {\cal E}_{\rm improved} + \parxx T^{0x}_{\rm improved} = 0
  \label{eq:32}
  \eeq
and for fields that decrease at large distances
  \beq
  E= \int \dd x {\cal E} = \int \dd x  {\cal E}_{\rm improved}\pp 
  \label{eq:33}
  \eeq
The improved energy density satisfies
  \beq
 {\cal E}_{\rm improved} = \fract12 T^{xx}
  \label{eq:34}
  \eeq
which is the nonrelativistic criterion of scale invariance in a field
theory.\cite{fifth} It follows from (\ref{eq:34}) that the dilation change
  \beq
 D = tE - \fract12 \int \dd{xx} {\cal P}
  \label{eq:35}
  \eeq
is time-independent:
  \beqar
   \frac{\dd{D}}{\dd{t}} &=& 
   E - \fract12 \int \dd{xx} \partt {\cal P}
   = E + \fract12 \int \dd{xx} \parxx T^{xx}\nonumber\\
  &=& \int\dd x \bigl({\cal E}_{\rm improved} - \fract12  T^{xx}
  \bigr) = 0 \pp 
  \label{eq:36}
  \eeqar

[Condition (\ref{eq:34}) also ensures conformal invariance in a
Galileo-invariant theory.\cite{fifth} Here the absence of Galileo
invariance requires absence of conformal invariance, as is seen from
the Lie algebra of the corresponding generators:
the bracket of $P$ with the conformal generator closes on the Galileo
boost generator.]

The momentum and energy of the soliton (\ref{eq:14}) are obtained
by evaluating $P$ and $E$ on the profile (\ref{eq:14}):
\beqar
P_{\rm soliton} &=& Mv \label{eq:37}\\[.5ex]
E_{\rm soliton} &=& \fract12 Mv^2 = \frac{1}{2M}P^2_{\rm soliton}
\label{eq:38}\\
M &=& mN\Bigl(1+\frac{\lambda^2}{12}N^2\Bigr) \label{eq:39}\\
\noalign{\noindent with}
 N &=& \int\dd x \rho =
\fract{2\hbar\alpha}{\lambda mv}\pp
\label{eq:40}
\eeqar
The soliton's dynamical characteristics are those of a nonrelativistic
particle of mass~$M$ moving with the velocity~$v$. The consequent
connection between energy and momentum exhibited by (\ref{eq:37})
and (\ref{eq:38}),
\beq
E_{\rm soliton} = \fract12 v P_{\rm soliton} 
\label{eq:41}
\eeq
in fact follows dilation invariance. For the soliton~(\ref{eq:14}), $\cal
P$ is a function $x-vt$, hence the time-independent dilation
change~(\ref{eq:35}) reads
\beqar
D &=& tE -\fract12\int\dd{xx} {\cal P}(x-vt)\nonumber\\
  &=& t\Bigl(E-\frac v2 P\Bigr) - \fract12 \int \dd{xx} {\cal P}(x)
\label{eq:42}
\eeqar
where the second equality follows from the first by a shift of
integration variable. Since both $D$ and the last term in (\ref{eq:41})
are time-independent, the coefficient of~$t$ must vanish, and this
implies~(\ref{eq:41}). Since for our solution ${\cal P}(x)$ is an even
function of~$x$, the last integral also vanishes and
\beq
D_{\rm soliton} = 0 \pp
\label{eq:43}
\eeq

\section{Other Solutions}
Thus far we have considered only the attractive interaction, which
binds a soliton: $g>0$ in (\ref{eq:1}) and $\lambda v>0$ in
(\ref{eq:13a}, \ref{eq:13b}). With a repulsive interaction, there also exist classical
solutions, variously known as ``kinks" or ``dark solitons".\cite{sixth}
They carry infinite energy, and we shall not discuss them here.

Unlike the conventional nonlinear Schr\"odinger equation~(\ref{eq:1}),
our chiral equation~(\ref{eq:8}) does not appear to be completely
integrable\,\cite{seventh}, and analytic expressions for multisoliton
solutions to~(\ref{eq:8}) are not available. It has been remarked,
however, that if the theory is modified by adding to~(\ref{eq:15}) the
potential
$V(\rho) = -\frac{\hbar^2\lambda^2}{8m}\rho^3$, then
Eq.\,(\ref{eq:16}) acquires the extra term
$-\frac{3\hbar^2\lambda^2}{8m}\rho^2\Psi$ and becomes an
integrable nonlinear, derivative Schr\"odinger equation with
nonlinearity
$i\frac{\hbar^2\lambda^2}{2m} \rho \parxx \Psi$. However, the
solitons are no longer chiral.\cite{eighth}

\section{Relation to (2+1)-dimensional Chern--Simons Theory}
Our dynamical, chiral model with Lagrange density (\ref{eq:15}) is
partially related, through dimensional reduction, to a
$(2+1)$-dimensional model of nonrelativistic fields interacting with a
$U(1)$ gauge potential, whose kinetic term is the Chern--Simons
expression.

Consider the $(2+1)$-dimensional Lagrange density
\beqar
{\cal L}_{(2+1)} &=&
\frac1{2\bar\kappa} \eps^{\alpha\beta\gamma} A_\alpha
\partial_\beta A_\gamma + 
i\hbar\Psi^* \Bigl(\partt + iA_0\Bigr)\Psi
-\frac{\hbar^2}{2m} \sum_{i=1}^2 \Bigr|\Bigl(\frac{\partial}{\partial
r^i} + i A_i\Bigr)\Psi\Bigr|^2\pp 
\label{eq:44}
\eeqar
When dependence on the second spatial coordinate is suppressed, and
$A_2$ is renamed $\frac{mc}{\hbar^2} B$, ${\cal L}_{(2+1)}$ leads to a
B--F gauge theory:
\beq
 {\cal L}_{(1+1)} = \frac1{2\kappa} B\eps^{\mu\nu} F_{\mu\nu}
+ i \hbar \Psi^*\Bigl(\partt + i A_0\Bigr)\Psi
  -\frac{\hbar^2}{2m}
\Bigr|\Bigl(\parxx + i A_x\Bigr)\Psi\Bigr|^2
-\frac{mc^2}{2\hbar^2} B^2 \rho \pp
\label{eq:44a}
\eeq
Here $F_{\mu\nu} = \frac{\partial}{\partial x^\mu} A_\nu -
\frac{\partial}{\partial x^\nu} A_\mu$, $x^\mu = \{t,x\}$, $A_\mu
= (A_0, A_x)$, and
$\kappa = \frac{\hbar^2}{mc}\bar\kappa$, where $c$ is the velocity
of light, which plays no role in the following. This is not yet our
theory~(\ref{eq:15}), because when the nonpropagating $B$ and
$A_\mu$ fields are eliminated, they decouple completely in the sense
that the phase of $\Psi$ may be adjusted so that all interactions
disappear:
\beq
{\cal L}_{(1+1)} \To i\hbar\Psi^*\partt \Psi - \frac{\hbar^2}{2m} 
\Bigl| \parxx \Psi\Bigr|^2 \pp
\label{eq:44b}
\eeq.

In order to make the vector potential $A_\mu$ and the $B$ field
dynamically active, thereby allowing the $\Psi$ particles to interact,
we include a kinetic term for $B$, which could be taken in the
Klein--Gordon form. However, we prefer a simpler expression that
describes ``chiral" Bose fields, propagating only in one direction, 
whose Lagrangian density is proportional to
 $\pm \frac{\partial B}{\partial t} \frac{\partial B}{\partial x} +
v\frac{\partial B}{\partial x} \frac{\partial B}{\partial x}$.\cite{ninth}
 Here $v$ is a velocity and the consequent equation of motion arising
from this kinetic term (without further interaction) is solved by
$B=B(x\mp vt)$ (with suitable boundary conditions at spatial infinity),
describing propagation in one direction, with velocity
$\pm v$.  Note that $\frac{\partial B}{\partial t} \frac{\partial
B}{\partial x}$ is not invariant against a Galileo transformation, which
is a symmetry of ${\cal L}_{(1+1)}$ and of $\frac{\partial B}{\partial
x} \frac{\partial B}{\partial x}$:  performing a Galileo boost on
$\frac{\partial B}{\partial t} \frac{\partial B}{\partial x}$ with
velocity $\tilde v$ gives rise to 
$\tilde v \frac{\partial B}{\partial x} \frac{\partial B}{\partial x}$,
 effectively boosting the $v$~parameter by
$\tilde v$. Consequently one can drop the $ v \frac{\partial B}{\partial
x} \frac{\partial B}{\partial x}$ contribution to the kinetic $B$
Lagrangian, thereby selecting to work in a global 
 ``rest frame". Boosting a solution in this rest frame then produces a
solution to the theory with a $\frac{\partial B}{\partial x}
\frac{\partial B}{\partial x}$ term. 

In view of the above, we choose the $B$-kinetic Lagrange density to be
   \beq
   {\cal L}_B = \pm \frac1{\hbar}\frac{\partial B}{\partial t} \frac{\partial B}{\partial x}
  \label{eq:45}
  \eeq
  and the total Lagrange density is $ {\cal L}_B + {\cal L}_{(1+1)}$. It is
possible to remove the $A_\mu$ and $B$ fields by a Hamiltonian
reduction, as described in Ref.\,\cite{tenth}, and by a phase
redefinition of $\Psi$. Once this is done, one is left with the Lagrange
density (\ref{eq:15})
  \beq
  {\cal L}_B + {\cal L}_{(1+1)} \To {\cal L}
  \label{eq:46}
  \eeq
with $\pm\kappa^2$ entering as $-\frac\lambda2$. The ``$\pm$" sign
reflects the sign arbitrariness of the  $\frac{\partial B}{\partial t}
\frac{\partial B}{\partial x}$ kinetic term, and without loss of
generality can be chosen so that $\lambda>0$. Also, the origin of
Galileo noninvariance is now recognized as arising from ${\cal
L}_B$.\cite{eleventh} 

\section{Quantum Theory}
The theory described by the Lagrange density (\ref{eq:15}) may be
quantized. For the quantum Hamiltonian we take the normal ordered
expression
\beq 
H =\frac{\hbar^2}{2m} \int\dd x : \Bigl|\Bigl(\parxx - i\frac\lambda2
\rho\Bigr)\Psi\Bigr|^2 : 
\label{eq:47}
\eeq 
and posit the canonical commutation relations~(\ref{eq:4}). 
It follows that the two-body wave function, defined as in~(\ref{eq:5}), satisfies
\beq
i\hbar\partt \psi_2 (t; x_1, x_2) =
-\frac{\hbar^2}{2m} \Bigl[ \frac{\partial^2}{\partial x_1^2} +
\frac{\partial^2}{\partial x_2^2} -
 i \lambda \delta(x_1-x_2)
\Bigl(\frac{\partial}{\partial x_1} +
\frac{\partial}{\partial x_2}\Bigr)\Bigr] \psi_2 (t; x_1, x_2)
\label{eq:48a}
\eeq 
Owing to time- and space-translation invariance, one can separate the time and
center-of-mass coordinates 
\beq
\psi_2(t; x_1, x_2) = e^{-i(Et/\hbar)} e^{i(P/\hbar)(x_1+x_2)/2} u(x_1-x_2) \pp
\label{eq:48b}
\eeq
The wave function for relative motion satisfies
\beq
\Bigl(-\frac{\hbar^2}m \frac{\partial^2}{\partial x^2} + \frac{P^2}{4m} - \hbar\frac
P{2m} \lambda\delta(x)\Bigr) u(x) = E u(x) \pp
\label{eq:48c}
\eeq
The presence of the total momentum $P$ 
in the $\delta$-function potential for relative motion
vividly demonstrates the absence of Galileo
invariance. Provided
\beq
\frac Pm>0
\label{eq:49}
\eeq
Eq.\,(\ref{eq:48c}) possesses a bound state solution with energy
\beq
 E=\frac{P^2}{4m} \Bigl(1-\frac{\lambda^2}{4}\Bigr)\pp
\label{eq:50}
\eeq
Since $P/m$ may be identified with (a multiple of) the classical
velocity $v$, we recognize the condition (\ref{eq:49}) as the same as
(\ref{eq:13b}), which guarantees soliton binding. So we expect that
there is a relation between the classical soliton and quantum bound
states. 

Justification for this can be see from the following argument. If we write (\ref{eq:50}) as
\beq
E = \frac{P^2}{2M_{\rm quantum}}
\label{eq:51}
\eeq 
we find that $M_{\rm quantum}$ is given by
\beq
M_{\rm quantum} = \frac{2m}{1-\frac{\lambda^2}{4}} \pp
\label{eq:52}
\eeq
Next we conjecture that the classical formula for $M$,
Eq.\,(\ref{eq:39}), should be modified in semiclassical quantization in
the same way as in the conventional nonlinear Schr\"odinger equation:
\emph{viz.}~as in (\ref{eq:6}) $N^3$ is diminished by~$N$.\cite{first}
Thus we replace (\ref{eq:39}) by
\beq
 M_{\rm semiclassical} = mN + \frac{m\lambda^2}{12} (N^3 - N) \pp
\label{eq:53}
\eeq
For $N=2$, this gives
\beq
 M_{\rm semiclassical} = 2m \Bigl(1+\frac{\lambda^2}4\Bigr) \pp
\label{eq:54}
\eeq
which agrees with (\ref{eq:52}) at weak coupling, {\it i.e.}, small
$\lambda^2$. Although explicit solutions to the $N$-body quantum
Schr\"odinger equation are not known for
$N>2$, one may establish perturbatively in $\lambda$ that
(\ref{eq:53}) is consistent with quantum bound states, at weak
coupling. 

The two-body problem inherits from the full  $N$-body field
theory the latter's symmetries and constants of motion. In particular,
the total momentum
   \beq
   P_2 \equiv p_1 + p_2
   \label{eq:67}
   \eeq
and the total energy
    \beqar
    H_2 &=& \frac{1}{2m} \bigl(p^2_1 + p^2_2\bigr) -
    \frac{\hbar}{2m}
    (p_1 + p_2)\lambda\delta(x_1-x_2) \nonumber\\
     &=& \frac{1}{2m} p^2 + \frac{1}{4m} P^2_2 -
    \frac{\hbar }{2m}P_2\lambda\delta(x) 
    \label{eq:68}\\
    &&
    \Bigl(p\equiv\fract12 (p_1-p_2),\quad x\equiv x_1-x_2\Bigr)
    \nonumber
    \eeqar
together with the dilaton charge
   \beqar
   D_2 &=& t H_2 - \fract14(x_1p_1 + p_1x_1 + x_2p_2 +
   p_2x_2)\nonumber\\
    &=& t H_2 - \fract14(XP_2 + P_2X + xp +
   px)
   \label{eq:69}\\
   &&\qquad\Bigl(X\equiv \fract12(x_1+x_2)\Bigr)\nonumber 
   \eeqar
are time-independent and close on the algebra\,\cite{fifth}
\beqar
{}[H_2,P_2] &=& 0 \label{eq:70a}\\
{}[H_2,D_2] &=& i\hbar H_2 \label{eq:70b}\\
{}[P_2,D_2] &=& i\hbar\fract12 P_2\pp \label{eq:70c}
\eeqar
It is interesting that in the present problem scale invariance does not
prevent the formation of a bound state. That is because the absence of
the Galileo invariance allows the bound state energy to depend on the
total momentum, which is not quantized, and as a consequence the
bound state energy (\ref{eq:50}) lies in the continuum, as is required
by scale invariance. 

\section{Conclusion} While our chiral, nonlinear Schr\"odinger
equation apparently cannot be completely integrated, it should be
possible, at least numerically, to study the $N$-soliton solution. From
this one could extract the nature of the solitons' mutual scattering and
possible dissipation. It should then be most interesting to examine the
same processes in the
$N$-body quantum Schr\"odinger equation. 
Also the procedure for quantizing the classical solutions could be
further developed.  While the analytic
intractability may be a daunting obstacle for arbitrary $N$, one
should at least unravel the $N=2$ case.


\begin{thebibliography}{11}
\bibitem{first}  C. Nohl, Ann Phys (NY) 96(1976), 234; P. Kulish, S.
Manakov, and L. Faddeev, Teor Mat Fiz 28(1976), 38 [English
translation Theor Math Phys 28(1976), 615].

\bibitem{second}  L. Dolan, Phys Rev D 13(1976), 528; Kulish \emph{et
al.}\    Ref.\,\cite{first}.

\bibitem{third} S.J. Benetton Rabello, 
 Phys Lett B363(1995),   180;  
Phys Rev Lett 76(1996), 4007, (E) 77(1996), 4851;
Stanford preprint SU-ITP\#96/11 (cond-mat/9604040). 
The reader must be alerted that errors mar these papers, and the
consequences drawn by the author are incorrect, as is discussed in the
\emph{Erratum} cited in Ref.\,\cite{first} and in Ref.\,\cite{fourth}.

\bibitem{fourth} The present review is based on U. Aglietti, L.
Griguolo, R. Jackiw, S.-Y. Pi, and D. Seminara, Phys Rev Lett 77(1996),
4406, and L. Griguolo and D. Seminara (in preparation). 

\bibitem{fifth} See, for example, R. Jackiw and S.-Y. Pi, Nucl Phys B
(Proc Suppl) 33C(1993), 104.

\bibitem{sixth} See Ref.\,\cite{fourth} and M. Toda, Nonlinear Waves
and Solitons  (Kluwer, Boston, 1989).

\bibitem{seventh} H.H.~Chen, Y.C.~Lee and C.S.~Liu,   Physica
Scripta  20(1979), 490; for a review see, for example, 
V.~Makhankov,   Soliton Phenomenology   (Kluwer, Boston, 1990),
ch.~2.

\bibitem{eighth} H. Min and Q-H. Park, Phys Lett B (in press).

\bibitem{ninth} R.~Floreanini and R.~Jackiw, Phys Rev Lett  59(1987),
1873. 

\bibitem{tenth} L.~Faddeev and R.~Jackiw, Phys Rev Lett   60(1988), 
1692. 

\bibitem{eleventh} In constrast to our direct dimensional reduction,
which is achieved by suppressing one coordinate, I.~Andri\'c,
V.~Bardek, and L.~Jonke (hep-th/9507110, August 1996) propose a
different, dynamically motivated reduction, which leads to the
Calogero--Sutherland model. 


\end{thebibliography}
\end{document}